\documentclass[9pt,preprint2,natbib209]{aastex}

\shorttitle{Convection and convective overshooting} \shortauthors{Jin, et al.}

\begin{document}


\title{Convection and convective overshooting in stars more massive than 10 $M_\odot$}
\author{Jie Jin\altaffilmark{1}, Chunhua Zhu\altaffilmark{1}, Guoliang L\"{u}\altaffilmark{1}}
\email{$^\dagger$978807464@qq.com, guolianglv@sina.com}
\altaffiltext{1}{School of Physical Science and Technology, Xinjiang
University, Urumqi, 830046, China.}





\begin{abstract}
In this paper,
four sets of evolutionary models are computed with
different values of the mixing length parameter
$\alpha_{\rm p}$ and the overshooting parameter
$\delta_{\rm ov}$.
The properties of the convective cores and
the convective envelopes are studied
in the massive stars.
We get three conclusions:
First, the larger $\alpha_{\rm p}$
leads to enhancing the convective mixing,
removing the chemical gradient,
and increasing the convective heat transfer efficiency.
Second, core potential
$\phi_{\rm c} = M_{\rm c} / R_{\rm c}$
describes sufficiently the evolution of a star, whether
it is a red or blue supergiant at central helium ignition.
Third, the discontinuity of hydrogen profile
above the hydrogen burning shell seriously
affect the occurrence of blue loops
in the Hertzsprung--Russell diagram.

\end{abstract}

\keywords{stars: evolution --- stars: interiors---stars: supergiants}

\section{INTRODUCTION}
Massive stars have convective cores and radiative
envelopes. Convection can mix different chemical elements
in the convection zone, and bring
heavy elements generated by nuclear reaction
region to the stellar surface.
It is also an important energy transport
mechanism.
Local mixing length theory (MLT) developed
by \cite{Vitense1958} is the most widely
used description for convection in stellar
model. The mixing length is proportional to
the pressure scale height
$l = \alpha_{\rm p} H_{\rm p}$, where
$\alpha_{\rm p}$ is a free parameter.
\cite{Pasetto2014} and \cite{Magic2014}
showed that the mixing length parameter $\alpha_{\rm p}$ was
of great importance in determining the convective
energy transport, and directly affected
the stellar interior structure of temperature.
However, from the theoretical point of view
the knowledge of $\alpha_{\rm p}$
is still a long standing problem.

Convective overshooting belongs to convection
and also affects energy transfer and mass
mixture in stars.
Classical treatment of the convective overshooting
is based on the non-local mixing length theory
\citep{Zahn1991}.
The predicted properties of stellar evolution are highly
sensitive to the degree of convective core overshooting
in the main sequence and core helium burning phases
\citep{Maeder1975,Maeder1976,Stothers1981,Bertelli1985,
Noels2010,Chen2002,Chen2003,Miglio2013}.
Moreover, \cite{Bressan2014} and \cite{Ding2014} showed
that convective overshooting and semiconvection
greatly affected the extent and the rate of mixing
in the outside deeply convective envelopes.

Red supergiants (RSGs) are evolved, massive
($10 - 30$ $M_{\rm \odot}$) core helium burning stars
\citep[e.g.,][]{Meynet2000,Eldridge2008,Brott2011,Davies2013}.
It represents a key phase in the evolution of massive stars.
However, convection overshooting in
the evolution of stellar models from blue to red in the post main
sequence phase and from red to blue in the blue loop phase
have not been studied systematically.

In this paper,
we attempt to investigate the properties
of the convective cores and the convective envelopes of
massive stars under different convection and
convective overshooting assumptions.
Special attention is given to the convective overshooting
at the bottom of convective envelope during the RSG phase.
\S 2 describes the evolution code, and gives the
input parameters for the modeling.
\S 3 gives the main results and discussion in detail.
In \S 4 we summarizes our conclusions.

\section{MODEL}
We employ an updated version of stellar evolution code ev,
which is based on the evolution program of
\cite{Eggleton1971,Eggleton1972,Eggleton1973}.
In this paper,
we use $499$ meshpoints in order to be more accurate.
Evolutionary models are computed for $10$, $15$, $20$, $30$,
$40$, $60$ $M_{\rm \odot}$ stars.
For each mass,
four sets of evolutionary models from the main sequence (MS)
to the end of core helium burning phase are computed (see Table $1$).
The first and third sets are computed without a certain amount of overshooting
and computed with radios $\alpha_{\rm p} = 1.5$ and $2.0$, respectively.
The second and fourth sets are computed with convective overshooting
$\delta_{\rm ov} = 0.12$ \citep{Pols1997,Pols1998}
and computed with radios $\alpha_{\rm p} = 1.5$ and $2.0$, respectively.
We neglect the effects of rotation, mass loss and magnetic fields.

\begin{table}[tbp]
\centering  
\begin{tabular}{lcc}  
\hline
&$\alpha_{\rm p}$ &$\delta_{\rm ov}$\\ \hline  
Case 1  &1.5  &0.0\\         
Case 2  &1.5  &0.12\\        
Case 3  &2.0  &0.0\\
Case 4  &2.0  &0.12\\ \hline
\end{tabular}
\caption{Four sets of evolutionary models.}
\end{table}

\section{RESULT}
\subsection{H-R diagram}
Figure $1$ shows
four cases of evolutionary tracks in the
Hertzsprung---Russell diagram (HRD) for each mass.
We can see that
the stars evolve toward lower effective temperature.
\cite{Eldridge2008} took $\mathrm{log} T_{\rm eff}\leq 3.66$
as a condition to be a RSG
during the central helium burning stage.
First we compare cases $1$ and $2$ for each mass.
It can be noted that
convective core overshooting
produces an extended track during
MS phase and
case $2$ has a larger luminosity
than case $1$.
\cite{Maeder1987} suggested that
models with an MS termination in the range of
spectral type $B0$ or later
($\mathrm{log} T_{\rm eff} < 4.47$),
the adopted convective overshooting
model produce an MS widening.
On one hand,
convective core overshooting makes the nuclear reaction region
larger, increases the MS lifetimes and
makes the chemical discontinuity displace farther from the stellar core.
On the other hand,
the extension of mixing increases more fresh hydrogen in
the convective core and enhances the
energy production.
The obtained conclusions are also applicable to
cases $3$ and $4$.

\begin{figure}[tbp]
\includegraphics[totalheight=3.0in,width=3.2in,angle=-90]{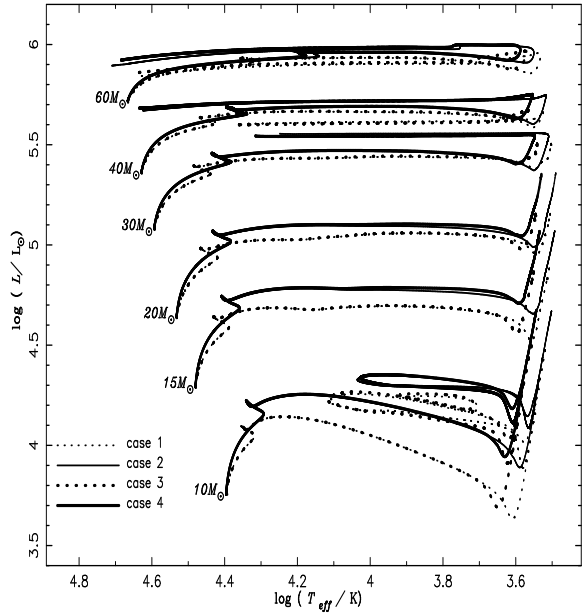}
\caption{Evolutionary tracks of 10, 15, 20, 30, 40,
and 60 $M_{\rm \odot}$ stars on the H-R diagram,
from the zero-age main sequence to
the exhaustion of central helium.}\label{fig:1}
\end{figure}

Second, we check cases $2$ and $4$ for each mass.
We find that case $4$ has a
higher effective temperature than case $2$
during the central helium burning phase.
It indicates that conducting heat of convection
is effected deeply by $\alpha_{\rm p}$.
In the view of fluid dynamics,
a larger value of $\alpha_{\rm p}$ means that
buoyancy can do more work on the convective cells.
Thus, the degree of convective mixing is enhanced,
and the efficiency of convective heat transfer
is increased.

\subsection{Internal structure}
\subsubsection{Lifetimes}
\begin{figure*}[tbp]
\begin{tabular}{cc}
\includegraphics[totalheight=3.0in,width=3.1in,angle=0]{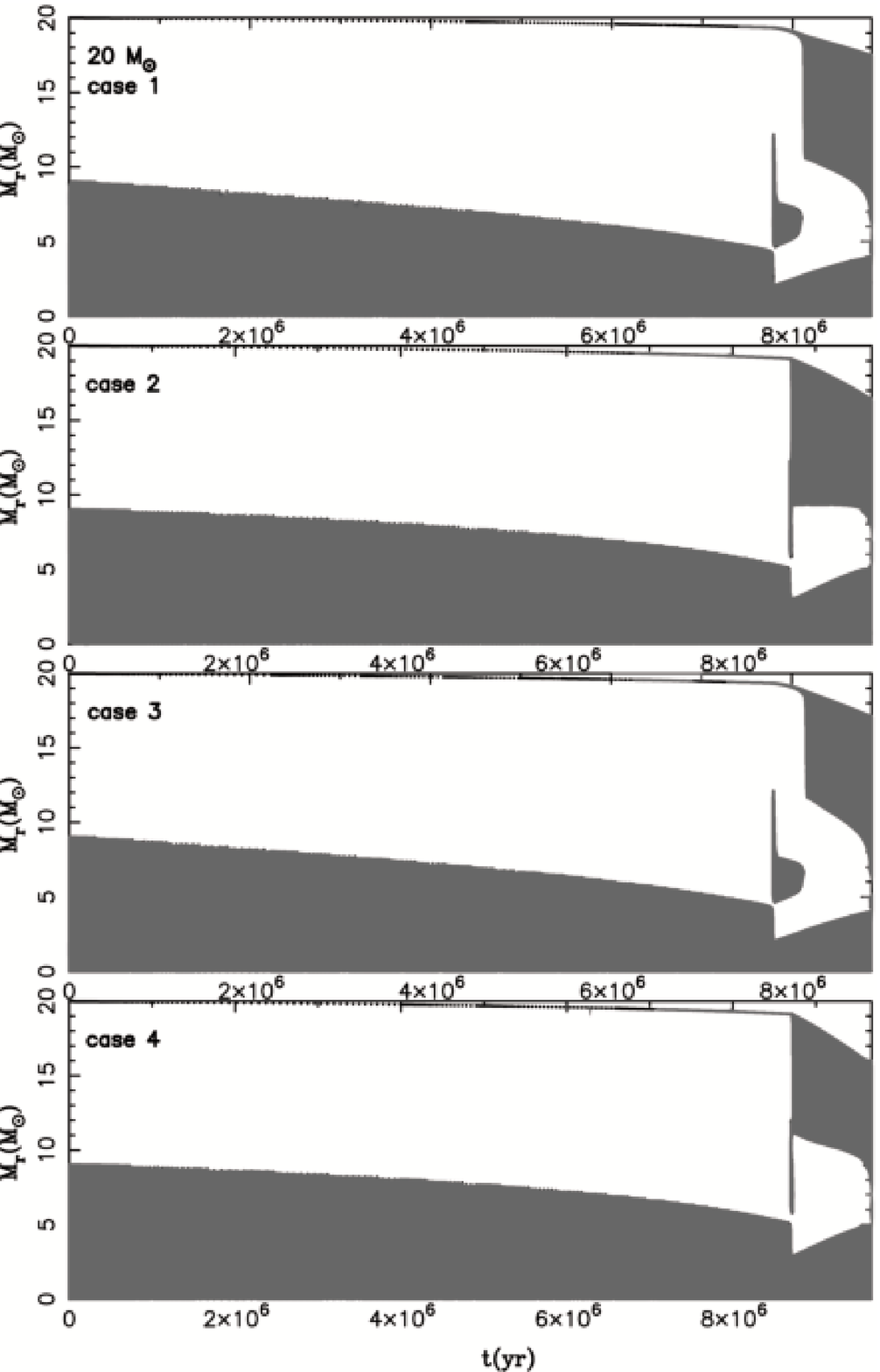}&
\includegraphics[totalheight=3.0in,width=3.1in,angle=0]{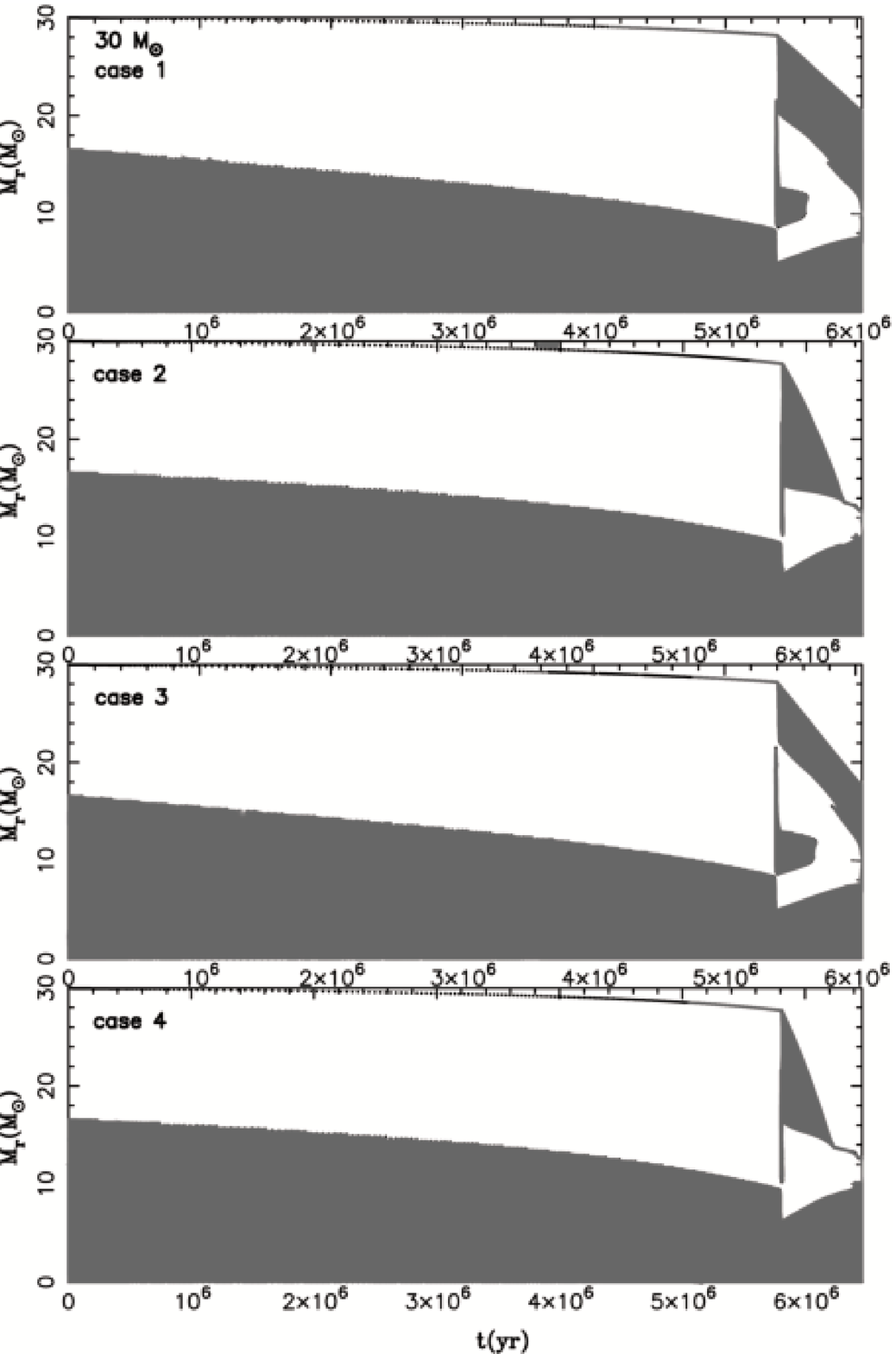}\\
\end{tabular}
\caption {The internal structure of $20$ and $30$ $M_{\rm \odot}$ stars,
which are from the MS to the exhaustion of central helium.
Filled regions represent the fully convective zones.}\label{fig:2}
\end{figure*}

Four cases of the internal structures of
$20$ and $30$ $M_{\rm \odot}$ stars
are showed in Figure $2$.
It is evident that core overshooting
significantly puts off the occurrence of the helium burning
phase and increases the MS lifetimes.
\cite{Maeder1987} found that
lifetime depended on the ratio of
$q_{\rm cc}M/L$,
where $q_{\rm cc}$ is the initial core mass
fraction, $\mathnormal{M}$ is stellar mass
and $\mathnormal{L}$ is stellar luminosity.
In cases $2$ and $4$,
there are more massive convective cores at the MS phase, and
the thermonuclear reactions increase the
MS lifetime $t_{\rm H}$.
However, during helium burning phase
the lifetime $t_{\rm He}$ is
decreased by convective core overshooting.
Due to the evolutionary tracks of models
still stay at a higher luminosity.
As a result, convective core overshooting reduces the
$t_{\rm He} / t_{\rm H}$ ratio.

\subsubsection{Convective envelope}
Our models show that during RSG phases
the stars possess deeply convective envelopes,
which locate in the superadiabatic layer at the top
rather than the hydrogen burning shell at the bottom.
\cite{Lai2012} indicated that
diffusive mixing in the
overshooting region would lead to the
increase of opacity and radiative temperature
gradient $\nabla_{\rm rad}$.
As a result,
a convective stable region below the
base of the convective envelope will be
converted to a convective unstable region.

It should be noticed that
case $1$ has a larger convective envelope than case $3$,
and case $2$ has a larger convective envelope than case $4$.
This indicates that
the larger $\alpha_{\rm p}$ produces more efficient
mixing in the chemical gradient region,
leads to a negligible chemical gradient
and suppresses the semiconvection.
\cite{Stothers1985} and \cite{Bressan1981} found that
core overshooting reduced
convective instability in the envelope.

It is striking to notice that
both cases $1$ and $3$ have a small intermediate
convective zone, just above the convective core,
in the early stage of core helium burning.
The reason is that the model without core overshooting leads
to a higher hydrogen abundance outside the convective core.
As a result, opacity and $\nabla_{\rm rad}$ will be larger.
So a semiconvection as well as full convection observed
in the early core helium burning phase.

\subsubsection{Stellar radius}
\begin{figure}[tbp]
\includegraphics[totalheight=3.0in,width=3.2in,angle=-90]{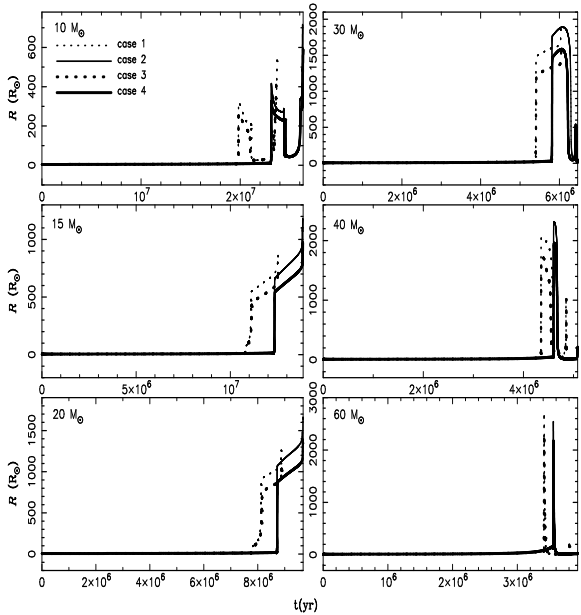}
\caption{The variations with time of the radius.}\label{fig:3}
\end{figure}

Figure $3$ shows the behaviour of radius
R (in solar units) as a function of time.
We can see that
case $1$ has a larger radius than case $3$, and
case $2$ has a larger radius than case $4$
during helium burning phase.
The radius of a star is defined as

\begin{equation}
\frac{R}{R_{\rm \odot}} =
\left( \frac{L}{L_{\rm \odot}}
\right) ^{1/2}
\left( \frac{T}{T_{\rm \odot}}
\right) ^{-2}
\end{equation}

Enhancing the efficiency of convection ($\alpha_{\rm p}$)
will increase the effective temperature of a star.
Hence radius decreases at a given luminosity.
\cite{Lai2012} also suggested that
the expansion of convection in the stellar envelope
increase convective heat transfer efficiency, and lead
to a smaller stellar radius and a higher effective temperature.

\subsubsection{Core mass}
\begin{figure}[tbp]
\includegraphics[totalheight=3.0in,width=3.2in,angle=-90]{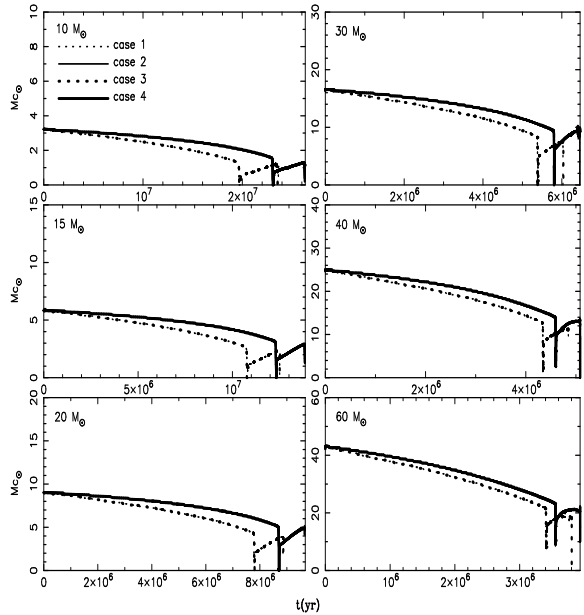}
\caption{The variations with time of the mass of central
convective hydrogen and helium core.}\label{fig:4}
\end{figure}

In figure $4$, for each mass the
evolution of the core mass as a function of
the time is presented.
It should be noticed that cases $1$ and $3$ overlap,
and cases $2$ and $4$ overlap.
It shows that the different value of $\alpha_{\rm p}$
have no effect on the mass of core.

\subsection{Core potential}

\begin{figure*}[tbp]
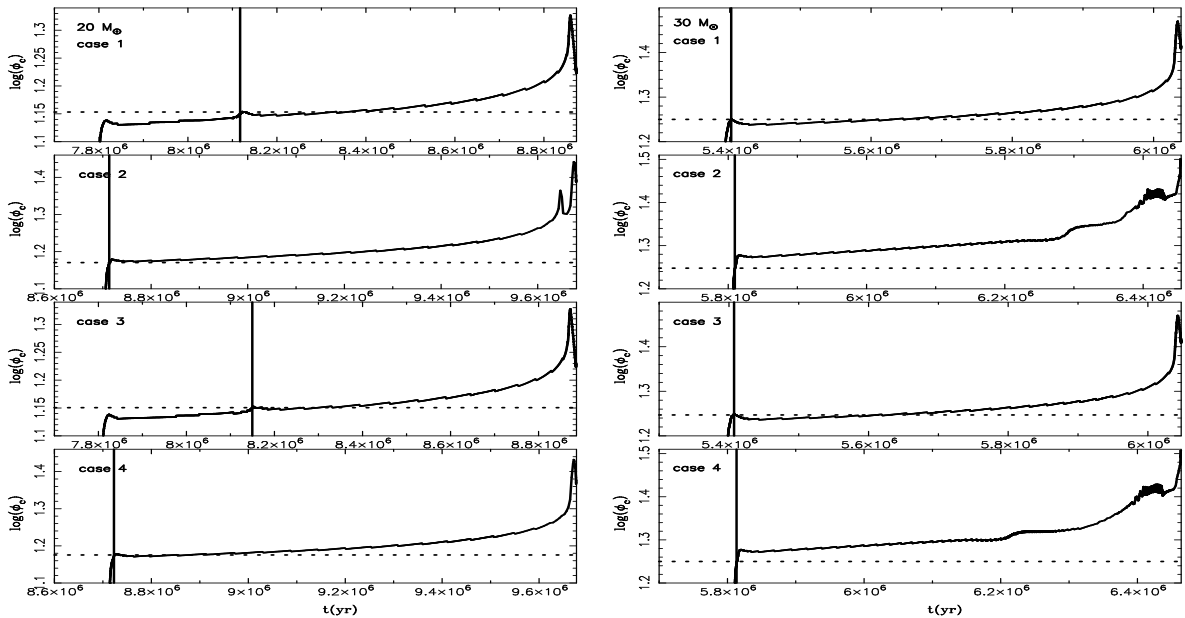

\begin{tabular}{cc}
\includegraphics[totalheight=3.0in,width=3.2in,angle=-90]{20shi.ps}&
\includegraphics[totalheight=3.0in,width=3.2in,angle=-90]{30shi.ps}\\
\end{tabular}
\caption {The variations with time of the core potential
in $20$ and $30$ $M_{\rm \odot}$ stars.
The dotted line represents the $\phi_{\rm 1cc}$.
The vertical full lines refer to the $\phi_{\rm 2cc}$.}\label{fig:5}
\end{figure*}

When does a massive star enter into RSG phase?
\cite{Lauterborn1971} suggested that the core potential
$\phi_{\rm c} = M_{\rm c} / R_{\rm c}$
describe sufficiently the evolution of stellar models
from blue to red in the post main sequence phase.
$M_{\rm c}$ and $R_{\rm c}$ are the mass
and radius (in solar units) of the core,
which extend from the centre up to
the base of the hydrogen burning shell.

We follow the work of \cite{Mowlavi1994}.
When a star closes to the Hayashi line,
a critical potential $\phi_{\rm 1cc}$
(defined by $\log(T_{\rm eff}) = 3.6$) can be found.
When $\phi_{\rm c}$ exceeds the critical value,
the star becomes a red (super)giant.
Furthermore, according to
the work of \cite{Eldridge2008},
we define the critical potential $\phi_{\rm 2cc}$
(defined by $\log(T_{\rm eff}) = 3.66$)
above which the star enter into RSG phase.
Figure $5$ shows the core potential changes with the time
in $20$ and $30$ $M_{\rm \odot}$ stars.

We can see that $\phi_{\rm 1cc}$ and
$\phi_{\rm 2cc}$ show a good agreement.
In cases $1$ and $3$, the $\phi_{\rm c}$ remain
a long time under the $\phi_{\rm cc}$.
The star is still in the
blue when central helium ignition.
The reason is that
the hydrogen burning shell grows rapidly
in a region of
smoothly varying hydrogen abundances.
When $\phi_{\rm c}$ reaches $\phi_{\rm cc}$,
the star becomes a red supergiant.
In cases $2$ and $4$,
$\phi_{\rm c}$ rapidly exceed $\phi_{\rm cc}$.
This result can be understand by considering
that the convective overshooting
makes a larger core
during the MS phase.
The star begins central helium burning as a
red supergiant.

\subsubsection{Blue loop}
However, when the hydrogen burning shell
encounters a discontinuity of hydrogen
profile during the central helium burning stage,
the issue will relate to the blue loop
\citep{Stothers1981,Stothers1991,Alongi1991}.
Our models develop huge convection envelopes
during RSG phases.
Downward convective overshooting at the bottom of
convective envelope will drive the hydrogen
discontinuity deeper into the interior.
When the hydrogen burning shell moves outward
and encounters the discontinuity,
fresh hydrogen will supply more nuclear fuels.
So the energy production is enhanced,
and the luminosity as well as the effective temperature
is increased.
The star then becomes a blue supergiant.
\cite{Lai2012} and \cite{Lauterborn1971}
showed that the occurrence of blue loop
critically depended on the hydrogen profile
just above the hydrogen burning shell.
However, \cite{Huang1983} and \cite{Bertelli1985} found that
convective core overshooting during MS
phase could significantly suppress the subsequent
development of the blue loop.
\cite{Xu2004} showed that convective core overshooting took
away the chemical gradient profile outside the stellar core, as
well as the abundance discontinuity formed later by
downward penetration of the convection envelope.

\section{Conclusions}
In this paper,
we investigate the properties of convective
cores and convective envelopes of
massive stars.
Different values of the overshooting parameter
$\delta$ and the mixing length parameter
$\alpha_{\rm p}$
result in the different theoretical predictions.
The main conclusions are summarized as follows:

1)Convective overshooting makes a larger
convective core, increases the MS lifetime
and extends MS band in the HRD.

2)Increasing the mixing length parameter
$\alpha_{\rm p}$ leads to enhancing
the convective mixing in the convective region
and increasing the convective heat transfer efficiency.
Thus, the star has a smaller chemical gradient,
a higher effective temperature
and a smaller stellar radius.

3)The core potential
$\phi_{\rm c} = M_{\rm c} / R_{\rm c}$
describes sufficiently the evolution of a star when
it moves from blue to red.
Without convective overshooting,
$\phi_{\rm c}$ slowly reaches $\phi_{\rm cc}$
and the star becomes a red (super)giant.
The reason is that the hydrogen burning shell grows
in a region of
smoothly varying hydrogen abundances.
With convective overshooting,
$\phi_{\rm c}$ rapidly exceed $\phi_{\rm cc}$
and then the star becomes a red (super)giant.
Because a larger core is produced
during the MS phase.

4)The discontinuity of hydrogen profile
above the hydrogen burning shell is
play an important role in
the occurrence of blue loops
in the HRD.

\section*{Acknowledgments}
This work was supported by
XinJiang Science Fund for Distinguished Young Scholars under Nos. 2013721014 and 2014721015,
the National Natural Science Foundation
of China under Nos. 11473024, 11363005 and 11163005, Foundation of Huoyingdong under No. 121107.

\bibliographystyle{apj}
\bibliography{jjapj}

\begin{thebibliography}{33}
\expandafter\ifx\csname natexlab\endcsname\relax\def\natexlab#1{#1}\fi

\bibitem[{{Alongi} {et~al.}(1991){Alongi}, {Bertelli}, {Bressan}, \&
  {Chiosi}}]{Alongi1991}
{Alongi}, M., {Bertelli}, G., {Bressan}, A., \& {Chiosi}, C. 1991, \aap, 244,
  95

\bibitem[{{Bertelli} {et~al.}(1985){Bertelli}, {Bressan}, \&
  {Chiosi}}]{Bertelli1985}
{Bertelli}, G., {Bressan}, A.~G., \& {Chiosi}, C. 1985, \aap, 150, 33

\bibitem[{{B{\"o}hm-Vitense}(1958)}]{Vitense1958}
{B{\"o}hm-Vitense}, E. 1958, \zap, 46, 108

\bibitem[{{Bressan} {et~al.}(2014){Bressan}, {Girardi}, {Marigo}, {Rosenfield},
  \& {Tang}}]{Bressan2014}
{Bressan}, A., {Girardi}, L., {Marigo}, P., {Rosenfield}, P., \& {Tang}, J.
  2014, ArXiv e-prints

\bibitem[{{Bressan} {et~al.}(1981){Bressan}, {Chiosi}, \&
  {Bertelli}}]{Bressan1981}
{Bressan}, A.~G., {Chiosi}, C., \& {Bertelli}, G. 1981, \aap, 102, 25

\bibitem[{{Brott} {et~al.}(2011){Brott}, {de Mink}, {Cantiello}, {Langer}, {de
  Koter}, {Evans}, {Hunter}, {Trundle}, \& {Vink}}]{Brott2011}
{Brott}, I., {et~al.} 2011, \aap, 530, A115

\bibitem[{{Chen} \& {Han}(2002)}]{Chen2002}
{Chen}, X., \& {Han}, Z. 2002, \mnras, 335, 948

\bibitem[{{Chen} \& {Han}(2003)}]{Chen2003}
---. 2003, \mnras, 341, 662

\bibitem[{{Davies} {et~al.}(2013){Davies}, {Kudritzki}, {Plez}, {Trager},
  {Lan{\c c}on}, {Gazak}, {Bergemann}, {Evans}, \& {Chiavassa}}]{Davies2013}
{Davies}, B., {et~al.} 2013, \apj, 767, 3

\bibitem[{{Ding} \& {Li}(2014)}]{Ding2014}
{Ding}, C.~Y., \& {Li}, Y. 2014, \mnras, 438, 1137

\bibitem[{{Eggleton}(1971)}]{Eggleton1971}
{Eggleton}, P.~P. 1971, \mnras, 151, 351

\bibitem[{{Eggleton}(1972)}]{Eggleton1972}
---. 1972, \mnras, 156, 361

\bibitem[{{Eggleton}(1973)}]{Eggleton1973}
---. 1973, \mnras, 163, 279

\bibitem[{{Eldridge} {et~al.}(2008){Eldridge}, {Izzard}, \&
  {Tout}}]{Eldridge2008}
{Eldridge}, J.~J., {Izzard}, R.~G., \& {Tout}, C.~A. 2008, \mnras, 384, 1109

\bibitem[{{Huang} \& {Weigert}(1983)}]{Huang1983}
{Huang}, R.~Q., \& {Weigert}, A. 1983, \aap, 127, 309

\bibitem[{{Lai} \& {Li}(2012)}]{Lai2012}
{Lai}, X.-J., \& {Li}, Y. 2012, ArXiv e-prints

\bibitem[{{Lauterborn} {et~al.}(1971){Lauterborn}, {Refsdal}, \&
  {Weigert}}]{Lauterborn1971}
{Lauterborn}, D., {Refsdal}, S., \& {Weigert}, A. 1971, \aap, 10, 97

\bibitem[{{Maeder}(1975)}]{Maeder1975}
{Maeder}, A. 1975, \aap, 43, 61

\bibitem[{{Maeder}(1976)}]{Maeder1976}
---. 1976, \aap, 47, 389

\bibitem[{{Maeder} \& {Meynet}(1987)}]{Maeder1987}
{Maeder}, A., \& {Meynet}, G. 1987, \aap, 182, 243

\bibitem[{{Magic} {et~al.}(2014){Magic}, {Weiss}, \& {Asplund}}]{Magic2014}
{Magic}, Z., {Weiss}, A., \& {Asplund}, M. 2014, ArXiv e-prints

\bibitem[{{Meynet} \& {Maeder}(2000)}]{Meynet2000}
{Meynet}, G., \& {Maeder}, A. 2000, \aap, 361, 101

\bibitem[{{Montalb{\'a}n} {et~al.}(2013){Montalb{\'a}n}, {Miglio}, {Noels},
  {Dupret}, {Scuflaire}, \& {Ventura}}]{Miglio2013}
{Montalb{\'a}n}, J., {Miglio}, A., {Noels}, A., {Dupret}, M.-A., {Scuflaire},
  R., \& {Ventura}, P. 2013, \apj, 766, 118

\bibitem[{{Mowlavi} \& {Forestini}(1994)}]{Mowlavi1994}
{Mowlavi}, N., \& {Forestini}, M. 1994, \aap, 282, 843

\bibitem[{{Noels} {et~al.}(2010){Noels}, {Montalban}, {Miglio}, {Godart}, \&
  {Ventura}}]{Noels2010}
{Noels}, A., {Montalban}, J., {Miglio}, A., {Godart}, M., \& {Ventura}, P.
  2010, \apss, 328, 227

\bibitem[{{Pasetto} {et~al.}(2014){Pasetto}, {Chiosi}, {Cropper}, \&
  {Grebel}}]{Pasetto2014}
{Pasetto}, S., {Chiosi}, C., {Cropper}, M., \& {Grebel}, E.~K. 2014, ArXiv
  e-prints

\bibitem[{{Pols} {et~al.}(1998){Pols}, {Schr{\"o}der}, {Hurley}, {Tout}, \&
  {Eggleton}}]{Pols1998}
{Pols}, O.~R., {Schr{\"o}der}, K.-P., {Hurley}, J.~R., {Tout}, C.~A., \&
  {Eggleton}, P.~P. 1998, \mnras, 298, 525

\bibitem[{{Schr{\"o}der} {et~al.}(1997){Schr{\"o}der}, {Pols}, \&
  {Eggleton}}]{Pols1997}
{Schr{\"o}der}, K.-P., {Pols}, O.~R., \& {Eggleton}, P.~P. 1997, \mnras, 285,
  696

\bibitem[{{Stothers} \& {Chin}(1981)}]{Stothers1981}
{Stothers}, R., \& {Chin}, C.-W. 1981, \apj, 247, 1063

\bibitem[{{Stothers} \& {Chin}(1985)}]{Stothers1985}
{Stothers}, R.~B., \& {Chin}, C.-W. 1985, \apj, 292, 222

\bibitem[{{Stothers} \& {Chin}(1991)}]{Stothers1991}
---. 1991, \apj, 374, 288

\bibitem[{{Xu} \& {Li}(2004)}]{Xu2004}
{Xu}, H.~Y., \& {Li}, Y. 2004, \aap, 418, 213

\bibitem[{{Zahn}(1991)}]{Zahn1991}
{Zahn}, J.-P. 1991, \aap, 252, 179

\end{thebibliography}


\label{lastpage}

\end{document}